\documentclass[reprint,superscriptaddress,amsmath,amssymb,aps,prb]{revtex4-2}

\usepackage{graphicx}% Include figure files

\begin{document}

\title{Entropic stress of grafted polymer chains in shear flow}

\author{Jan Mees}
\affiliation{Department of Microsystems Engineering, University of Freiburg, Georges-K\"ohler-Allee 103, 79110 Freiburg, Germany}
\affiliation{Cluster of Excellence livMatS, Freiburg Center for Interactive Materials and Bioinspired Technologies, University of Freiburg, Georges-K\"ohler-Allee 105, 79110 Freiburg, Germany}
\author{Thomas C. O'Connor}
\affiliation{Department of Materials Science and Engineering, Carnegie Mellon University, Pittsburgh, Pennsylvania 15213, USA}
\author{Lars Pastewka}%
 \email{lars.pastewka@imtek.uni-freiburg.de}
\affiliation{Department of Microsystems Engineering, University of Freiburg, Georges-K\"ohler-Allee 103, 79110 Freiburg, Germany}
\affiliation{Cluster of Excellence livMatS, Freiburg Center for Interactive Materials and Bioinspired Technologies, University of Freiburg, Georges-K\"ohler-Allee 105, 79110 Freiburg, Germany}

\date{\today}

\begin{abstract}
We analyze the shear response of grafted polymer chains in shear flow via coarse-grained molecular dynamics simulations. Our simulations confirm that the shear response is dominated by the brush's outermost correlation volume, which depends on shear rate at high Weissenberg number. The system's shear stress can be approximated by the brush's entropic stress. The simulations further reveal that at low Weissenberg number, the entropic shear stress of grafted chains is independent of the Weissenberg number. Increasing the Weissenberg number leads to Wi-dependent behavior: chains first reorient along the shear direction and elongate at higher Wi. The entropic shear stress increases linearly with Weissenberg number in this regime. We relate these calculations to experimental observations on the velocity dependence of brush and hydrogel friction.
\end{abstract}

\maketitle

\section{Introduction}

Polymer brushes are dense systems of grafted polymer chains. They have been used in varied applications such as stabilizing colloidal suspensions~\cite{napper1983,Gast1986M,Borukhov2000PRE}, anti-fouling coatings~\cite{Hucknall2009Biointerphases,Xu2016ACSAMI} and adjusting surface interactions and wetting behavior~\cite{Mansky1997Science}. Brushes are known for their excellent lubricating properties~\cite{Klein1996ARMS,Binder2011SM,Kreer2016SM}, as are related soft systems of biological importance such as hydrogels~\cite{Gong2006SM,Dunn2014TRIL,Pitenis2014SM}. 

For solvated brushes, there is an ongoing debate over the roles of solvent hydrodynamics and nonequilibrium polymer conformations in setting the experimentally observed frictional behavior~\cite{Pitenis2014SM}. 
In this paper we apply molecular simulations to study the contribution of grafted polymer's to the interfacial shear stress of solvated brushes. We model systems of grafted chains with varying chain length and grafting density (see Fig.~\ref{fgr:grafted_chains}a and b), and analyze their microscopic response to a solvent shear flow. 
We focus on the case where the brush surface is far from any other solid surfaces, complementing our recent study on friction between contacting polymer interfaces~\cite{Mees2022arXiv}.

At low grafting densities, in what is called the ``mushroom'' regime, chains do not interact with each other. With increasing grafting density $\Gamma$, chains start overlapping and adopt increasingly straight, brush-like conformations due to excluded volume effects~\cite{Binder2011SM}. The transition between both regimes occurs at $\Gamma\approx\Gamma^*$, with $\Gamma^*=1/(\pi R_\text{g}^2)$ where $R_\text{g}$ is the radius of gyration of the individual chain inside the brush.

When exposed to shear flow, the solvent's velocity penetrates a certain distance into the brush before it is screened out~\cite{Milner1991M}. This momentum transport into the brush leads to tilting of chains and stretching in the direction of shear~\cite{Lai1993JCP,Peters1995PRE,Miao1996M,Grest1996PRL,Doyle1997PRL,Doyle1998M,grest1999}. As a result,  a grafted polymer chain in shear flow is subject to three forces: an osmotic force due to the gradient in polymer concentration across the interface, a shear force due to the solvent's shear flow and a restoring force due to conformational entropy~\cite{Rabin1990EPL}. 
Studies have shown that these forces compete such that, at any given time, only a fraction of chains will be exposed to the shear flow, while the other chains remain inside the brush, cycling between both states~\cite{Aubouy1996JdPII,grest1999,Clement2001EPL,Muller2007EPL}. 

Several theoretical models have been proposed for polymer brushes exposed to shear flow~\cite{Rabin1990EPL,Barrat1992M,Kumaran1993M,Harden1996PRE,Aubouy1996JdPII,Clement2001EPL}. The state-of-the-art model by Clement et al.~\cite{Clement2001EPL} builds on the fact that shear flow only penetrates over the first correlation length $\xi$ into the dense brush. The system's shear response is determined by the competition between the shear rate $\dot{\gamma}$ and the lateral relaxation time  $\tau_\text{r}$ of the chains outermost "blob" or correlation volume $\xi^3$. The system's response can then be characterized by the Deborah number $\text{De}=\dot{\gamma}\tau_\text{r}$, which in this case is mathematically equivalent to the Weissenberg number Wi~\cite{Pitenis2018TRIL}. Assuming that within each blob, the chain segments follow Zimm dynamics~\cite{Harden1996PRE,Aubouy1996JdPII,Clement2001EPL}, the relaxation time is given by~\cite{rubinstein2003}
\begin{equation}
    \tau_\text{r}\approx \eta_\text{s} \xi^3/k_\text{B}T,
    \label{eq:zimm}
\end{equation}
where $\eta_\text{s}$ is the solvent viscosity.

In this work, we explore the polymer's contribution to the shear stress by computing the conformational entropy (entropic stress) of grafted chains in shear flow~\cite{doi1988,OConnor2018PRL}. This allows us to disentangle the polymer's contribution to the shear stress from the solvent's viscous dissipation. Following recent conventions in hydrogel and polymer brush friction, we characterize our system's response as a function of the Weissenberg number Wi~\cite{Galuschko2010Langmuir,Spirin2010EPJE,Uruena2015Biotribology,Pitenis2018TRIL}.

We show that brush's outermost correlation volume determines the brush's shear response and how its size varies with shear rate and grafting density, information which is not experimentally accessible. We further show that the conformational entropy of grafted chains fully explains the system's shear stress and friction in semi-dilute and dense brushes. This entropic shear stress is independent of Wi at low Wi, where chains can relax faster than they are deformed. The entropic stress increases linearly after a threshold Wi, in a regime associated with chain reorientation in the direction of shear. 
At higher Wi, greater than the Rouse time, grafted chains elongate but the entropic stress still depends linearly on Wi.

Crucially, our results indicate that these results depend on detail of the definition of Wi.
In the regime where chains elongate, the entropic stress appears to depend sublinearly on Wi if a shear-rate-independent relaxation time is used to compute the Weissenberg number, as is typical for experiments.
In our simulations, this analysis leads to power-law exponents similar to what has been found in polymer brush and self-mated hydrogel friction~\cite{Galuschko2010Langmuir,Uruena2015Biotribology,Pitenis2018TRIL}.

\section{Theory}

The stress tensor in a polymer solution is comprised of contributions by the solvent's viscous dissipation, the hydrostatic pressure and polymer effects. The stress tensor is given by~\cite{doi1988}
\begin{equation}
    \sigma_{\alpha\beta}=\eta_\text{s} \left(\kappa_{\alpha\beta}+\kappa_{\beta\alpha}\right) - p\delta_{\alpha\beta} +\frac{\rho}{n} \langle f_{\alpha}r_{\beta}\rangle,
\end{equation}
where $\alpha$ and $\beta$ denote Cartesian coordinates. The solvent's viscous dissipation yields a contribution $\eta_\text{s}(\kappa_{\alpha\beta}+\kappa_{\beta\alpha})$, where $\kappa_{\alpha\beta}$ is the local shear rate (or velocity gradient). The hydrostatic pressure's contribution is $-p\delta_{\alpha\beta}$, with  pressure $p $ and  Kronecker delta $\delta_{\alpha\beta}$. $\langle \cdot \rangle$ represents the ensemble average. Finally, the polymer's contribution to the stress tensor is 
\begin{equation}
    \sigma_{\alpha\beta}^\text{entr}=\rho \langle f_{\alpha}r_{\beta}\rangle/n.
    \label{eq:polymer-stress}
\end{equation}
Here $\rho$ is the polymer's number density and $n$ is the number of bonds per chain. The force $\vec{f}$ is the sum of non-hydrodynamic forces acting on a given chain, and $\vec{r}$ is the chain's end-to-end distance vector. 
 
The polymer's contribution to the stress tensor is given by the equation for the non-linear, entropic force of a polymer chain $|\vec{f}|  = k_{\rm{B}}T \mathcal{L}^{-1}(h)/l_{\rm{k}}$~\cite{OConnor2018PRL}. Here $l_{\rm{k}}$ is the Kuhn length, $\mathcal{L}^{-1}$ is the inverse Langevin function~\cite{rubinstein2003} and $h=|\vec{r}|/n b$ is the extension ratio with bond length $b$. The extension ratio is approximately zero for an ideal, Gaussian chain and unity for a fully extended chain. The Kuhn length is defined as $l_{\rm{k}}=C_\infty b$, and $C_\infty$ is Flory's characteristic ratio. From Eq.~\eqref{eq:polymer-stress} we obtain
\begin{equation}\label{eq:entropicstress}
    \sigma_{\alpha\beta}^{\rm{entr}} = \frac{\rho(\dot{\gamma}) k_{\rm{B}}T }{C_\infty} \langle \mathcal{L}^{-1}\left(h\right) h~\hat{r}_\alpha \hat{r}_\beta \rangle,
\end{equation}
where $\hat{r}_\alpha$ are components of the unit vector 
$\vec{r}/r$ of the end-to-end distance. Flory's characteristic ratio is $C_\infty=1.7$ for our system (see Appendix for further details). The number density is given by $\rho=N_\text{m}/V$, where $N_\text{m}$ is the number of monomers and $V=L_x L_y H(\dot{\gamma})$ is the volume occupied by the grafted chains, with $H$ being the brush's height. The number density is shear rate-dependent because at higher shear rates the chains tilt and stretch, changing the volume.

This derivation ignores changes to the chain's internal energy which might arise at large extensions, and hence only captures the effects of changes in configurational entropy, which is usually the dominant source of stress in shearing flows. For the purpose of clarity and brevity, we henceforth call the polymer's contribution to the stress tensor the entropic stress $\sigma^\text{entr}_{\alpha\beta}$. Two different molecular effects go into the entropic stress. First, the hardening factor  $\mathcal{L}^{-1}\left(h\right)h$ captures the mechanical effect of chain elongation. Second, the orientation factor $\hat{r}_\alpha \hat{r}_\beta$ captures the alignment of polymer chains.

\section{Methods}
\begin{figure*}
\centering
  \includegraphics{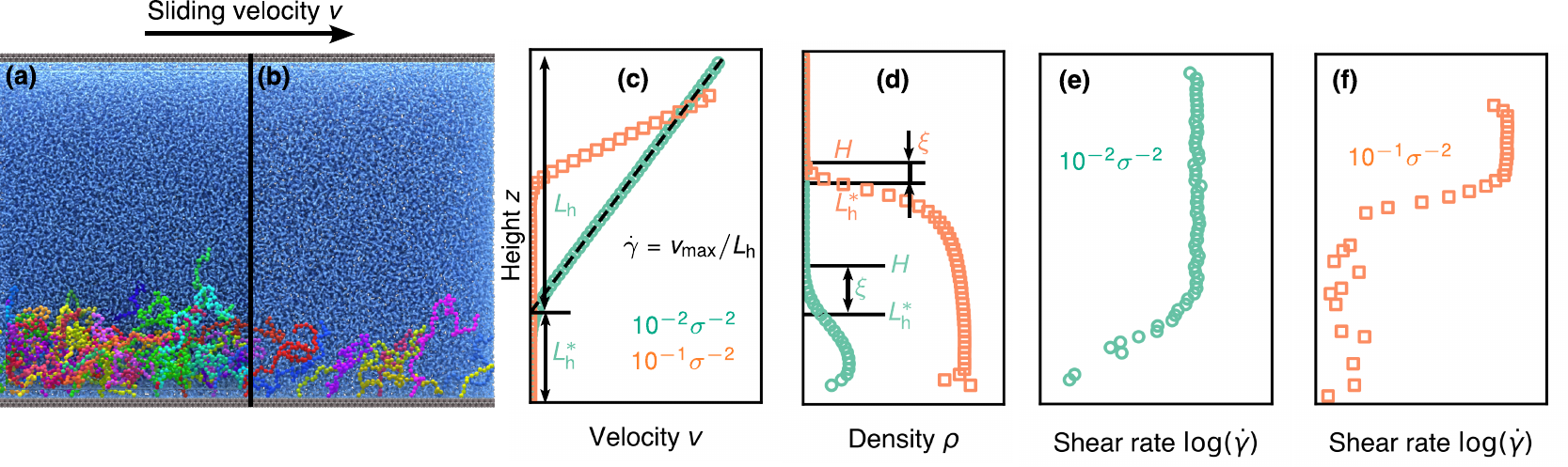}
  \caption{(a) Polymer chains with 60~beads/chain and grafting density $\Gamma=10^{-2}~\sigma^{-2}$ at sliding speed $v=5\cdot10^{-2}~\sigma \tau^{-1}$. (b) Only two rows of grafted chains are depicted for clarity. Each chain has a different color to ease visibility. (c) The solvent's velocity profile in the shear direction is used to determine the effective shear rate $\dot{\gamma}$, here shown for a system with $N=60$~beads/chain at grafting densities $\Gamma=10^{-2}~\sigma^{-2}$ (green) and $10^{-1}~\sigma^{-2}$ (orange). Dashed lines represent fits. (d) Polymer density distribution as a function of height for the same systems. The size of the outermost blob $\xi$ is determined as the difference of brush height $H$ and $L^*_\text{h}$ (e) and (f) show the height dependency of the local shear rate $\dot{\gamma}(z)=\Delta v/\Delta z$. }
  \label{fgr:grafted_chains}
\end{figure*}

Our molecular system, depicted in Figure~\ref{fgr:grafted_chains}a and b, consists of polymer chains bonded to a face-centered cubic wall with lattice spacing $1.2\,\sigma$ and the [111] crystallographic direction facing the interface.  The chains are bonded to the exposed wall beads following an hexagonal pattern with grafting density $\Gamma$.
The polymers are described by a coarse-grained bead-spring model~\cite{Kremer1990JCP} in which the excluded volume is modeled by a Lennard-Jones (LJ) potential with interaction energy $\varepsilon=1.0$, length $\sigma=1.0$ and cut-off radius $r_{\mathrm{cut}}=1.6~\sigma$. Masses are set to unity, so that the time-scale of the simulation is given by $\tau=\sigma (m/\varepsilon)^{1/2}$.  Consecutive monomers (polymer beads) are connected by a nonlinear FENE bond with spring constant $k=30.0~\varepsilon \sigma^{-2}$ and maximum extension $R_0=1.5~\sigma$~\cite{Armstrong1974JCP,Kremer1990JCP}. The cut-off radius of the LJ interaction of bonded monomers is set to $r_{\mathrm{cut}}=2^{1/6}~\sigma$, making the LJ interaction purely repulsive. Monomers interact with wall beads via an LJ interaction with $\varepsilon=0.6$, $\sigma=1.3$ and $r_{\text{cut}}=1.6~\sigma$. The monomers are bonded to the wall via FENE bonds with spring constant $k=30.0~\varepsilon \sigma^{-2}$ and maximum extension $R_0=2.1~\sigma$. The LJ interaction energy is set to $\varepsilon=0.8$, the length is set to $\sigma=1.5$ and the cut-off radius to $r_{\text{cut}}=2^{1/6}~\sigma$. 

We explicitly model a dimer solvent made of chains of length $N=2$~\cite{Grest1997COCIS,grest1997dynamics}.
This type of explicit solvent inhibits layering effects at the walls~\cite{Galuschko2010Langmuir}. The dimers are connected by FENE bonds with $k=30.0~\varepsilon \sigma^{-2}$ and maximum extension $R_0=1.3~\sigma$. The LJ  parameters are $\varepsilon=0.8$, $\sigma=0.8$ and $r_{\text{cut}}=2^{1/6}~\sigma$. Non-bonded dimers  interact via an LJ potential with parameters $\varepsilon=0.5$, $\sigma=1.0$ and $r_{\text{cut}}=2.5~\sigma$. The parameters for dimer-polymer interactions are set to $\varepsilon=1.2$, $\sigma=1.0$ and $r_{\text{cut}}=2.5~\sigma$. Finally, dimers interact with the wall with $\varepsilon=0.6$, $\sigma=1.0$ and $r_{\text{cut}}=2.5~\sigma$. The aforementioned parameters are identical to the parameters used by de Beer et al.~\cite{De2013SM} to induce good-solvent conditions.
The combined polymer and solvent system is simulated at a fixed concentration of $0.8$~beads~$\sigma^{-3}$. A representative system has the dimensions $L_x=100~\sigma,~L_y= 87~\sigma,~ L_z=64~\sigma$ and contains 465632~beads. 

We use a DPD thermostat with friction coefficient $\gamma=5~m \tau^{-1}$ and cutoff $r_{\rm c}=2^{1/6}~\sigma$ to thermalize the solvent and the polymers to a temperature of $k_\text{B} T=0.6~\varepsilon$~\cite{Espanol1995EPL,Hoogerbrugge1992EPL}, where $k_\text{B}$ is the Boltzmann constant. The top wall is moved at constant sliding speed $v$ along the $x$ axis in order to induce a shear flow in the solvent. A downward pressure of $5\cdot10^{-4}~m \sigma^{-1}~\tau^{-2}$ acts on the top wall, slightly compressing the system. We use a time-step of $\Delta t=5\cdot10^{-3}~\tau$ in our simulations.

\section{Results}

\subsection{Shear stress}

By comparing the solvent's velocity profile to the polymer's density distribution for systems with $N=60$~beads/chain (Figure~\ref{fgr:grafted_chains}c and d), we observe that within the brush, hydrodynamic interactions are screened by the polymer and the solvent's velocity decreases rapidly inside the brush. Above the brush, the solvent acts as a Newtonian fluid and a Couette flow is established. 

The brush itself is exposed to the shearing solvent.
The key parameter is therefore the shear rate $\dot{\gamma}=v/L_\text{h}$ of the solvent, and not the sliding velocity $v$.
Here $L_\text{h}$ is the length of the solvent's Couette region, 
which we compute by linearly extrapolating the (linear) velocity profile inside the solvent.
The overall simulation cell is of height $L_\text{h} + L_\text{h}^*$ (see Fig.~\ref{fgr:grafted_chains}c), where $L_\text{h}^*$ is the height of the quiescent part of the brush.
While we determine $\dot{\gamma}$ from our simulations directly at high velocities, the shear rates obtained at low sliding velocity are dominated by thermal fluctuations.
For sliding speeds $v<v'$, we therefore compute $\dot{\gamma}$ by assuming that $L_\text{h}$ does not change from its value at $v'=5\cdot10^{-3}~\sigma\tau^{-1}$.

We analyze the shear response as a function of the Weissenberg number $\text{Wi}=\dot{\gamma} \tau_\text{r}$.
As argued in the introduction, $\tau_\text{r}$ is the polymer relaxation time $\tau_\mathrm{r}$, Eq.~\eqref{eq:zimm}, of the outermost correlation volume of size $\xi^3$.
The corresponding correlation length $\xi$ is identical to the hydrodynamic penetration~\cite{Harden1996PRE,Clement2001EPL}, which we obtain from the difference between the brush's height $H$ and the end of the Couette flow region $L_\text{h}^*$, $\xi=H-L_\text{h}^*$ (Fig.~\ref{fgr:grafted_chains}c and d).
We compute the brush height $H$ as in Ref.~\cite{Vos2009Polymer}, from the position at which the density has decayed to 1~\% of its maximum value.
Since $\xi$ changes with shear rate, this implies that $\tau_\text{r}$ also depends on shear rate through Eq.~\eqref{eq:zimm}.

The influence of the grafting density on the velocity profile can be observed in the logarithmic depiction of the shear rate (Fig.~\ref{fgr:grafted_chains}e and f). For $\Gamma=10^{-2}~\sigma^{-2}$, the shear rate gradually increases to its maximum value $\dot{\gamma}_0$, while the transition is much sharper for $\Gamma=10^{-1}~\sigma^{-2}$. As a reference, for systems with $N=60~\text{beads/chain}$ the grafting density at which chains begin to overlap is $\Gamma^*=1/(\pi R_\text{g}^2) \approx9\cdot10^{-3}~\sigma^{-2}$. %The rate with which the solvent's velocity decays inside the brush can be quantified by fitting the shear rate to a Fermi-Dirac-like equation 
% Shear rate sharpness of transition, fit equation, screening length
%\begin{equation}\label{eq:shearrate}
%    \dot{\gamma}(z) = \dot{\gamma}_0\left[1+e^{(H-z)/\lambda}\right]^{-1},
%\end{equation}
%where $\dot{\gamma}_0$ is the maximum shear rate, $H$ is the brush height and $\lambda$ is a screening length.
%\lp{LP2JM: H is reused here but different from above. Check whether this can be consolidated.}
%
%Fitting all three parameters to the simulated velocity profiles yields a screening length that is smaller ($\lambda=1.0~\sigma$) for the more densely grafted system than for the system with lower grafting density ($\lambda=2.4~\sigma$).

\begin{figure}
\centering
  \includegraphics{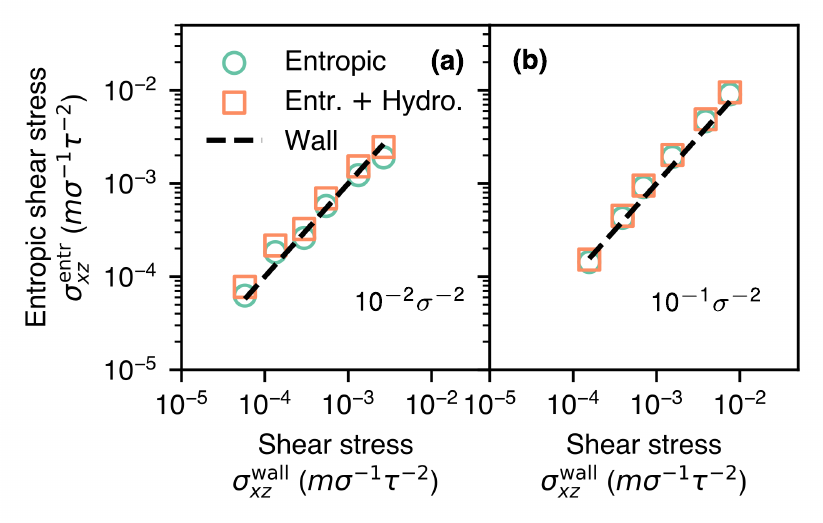}
  \caption{Entropic and hydrodynamic shear stress as a function of the system's shear stress measured at the walls at grafting densities (a) $\Gamma=10^{-2}~\sigma^{-2}$ and (b) $\Gamma=10^{-1}~\sigma^{-2}$ for a system with 60~beads/chain. Each measurement corresponds to a different sliding velocity. Dashed line represents system's shear stress.}
  \label{fgr:walls}
\end{figure}

Mechanical equilibrium in our simulation geometry means that the shear stress is constant along the simulation cell's height ($\partial \sigma_{xz}/\partial z =0$). It follows that the the system's shear stress measured at the walls $\sigma_{xz}^{\mathrm{wall}}$ is equal to the Newtonian, viscous shear stress in the Couette-like region, $\sigma^{\text{visc}}_{xz}=\eta_\text{s}\dot{\gamma}$.
Within the brush, the shear stress will be comprised of the sum of the entropic stress and a hydrodynamic, viscous stress~\cite{Harden1996PRE}. This decomposition is shown in Fig.~\ref{fgr:walls}a, where the entropic stress is obtained from Eq.~\eqref{eq:entropicstress}.
The solvent's hydrodynamic contribution to the shear stress is given by the average shear rate inside the brush 
\begin{equation}
    \sigma_{xz}^{\text{hydro}}=\frac{\eta_\text{s}}{H} \int_0^H \text{d}z\, \dot{\gamma}(z).
\end{equation}
Figure~\ref{fgr:walls}a shows that the entropic stress is approximately equal to the system's shear stress for grafting densities $\Gamma=10^{-2}~\sigma^{-2}$ and $\Gamma=10^{-1}~\sigma^{-2}$ (Fig.~\ref{fgr:walls}). 
In both cases, the hydrodynamic stress within the brush is much smaller than the entropic stress.
%
%This shows that the overall stress can be additively decomposed in entropic and viscous (solvent) contributions, but that it can be approximated by the entropic stress. 
%

\subsection{Choice of relaxation time}

Next, we study the influence of the choice of relaxation time in the definition of the Weissenberg number. The polymer relaxation time is often times assumed as being a constant when computing the Weissenberg number to evaluate experimental and simulation data in applications such as polymer brush friction~\cite{Galuschko2010Langmuir,Spirin2010EPJE} or self-mated hydrogel friction~\cite{Uruena2015Biotribology,Pitenis2018TRIL,Mees2022arXiv}. It then compares an equilibrium measure (the relaxation time) with a non-equilibrium measure (the shear rate).
In this paper, we denoted this traditional definition of the Weissenberg number by $\text{Wi}^*= \dot{\gamma} \tau^*_\text{r}$, and the corresponding equilibrium relaxation time of a single, dilute chain of length $N$ by $\tau_\text{r}^*$. Note that $\tau_\text{r}^*$ does not depend on shear rate $\dot{\gamma}$ while the true relaxation time $\tau_\text{r}$ does. The implications of this are discussed further below.

The relaxation time is computed in molecular dynamics simulations via the time-autocorrelation function of the trace of the gyration tensor $\langle\sum_\alpha R^2_{\text{g},\alpha\alpha}\rangle \propto \exp(-t/\tau^*_\text{r})$~\cite{doi1988}, rather than through an analytical estimate such as Eq.~\eqref{eq:zimm}. We have confirmed that the relaxation time computed from the analytical estimate, Eq.~\ref{eq:zimm}, is close (within a factor of two) to the relaxation time obtained from single-chain simulations. 

\begin{figure}
\centering
  \includegraphics{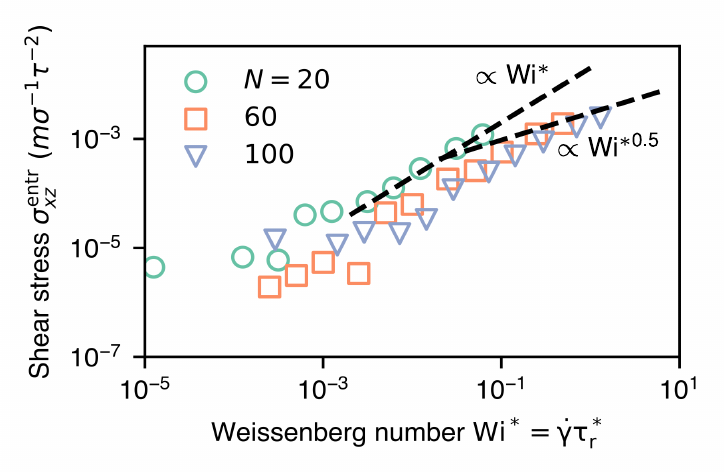}
  \caption{Entropic shear stress as a function of chain length and
  Weissenberg number for $\Gamma=10^{-2}~\sigma^{-2}$, using the relaxation time  $\tau^*_\text{r}$ of dilute chains of length $N$.}
  \label{fgr:relaxation}
\end{figure}

Figure~\ref{fgr:relaxation} shows the entropic stress as a function of $\text{Wi}^*$.
There is a $\text{Wi}^*$-independent regime at low $\text{Wi}^*$, followed by a linear regime which crosses over to a sublinear regime with an exponent of approximately $0.5$ at high $\text{Wi}^*$ (Fig.~\ref{fgr:relaxation}).
The linear regime coincides with an increase in the orientation factor, while in the sublinear regime the orientation factor remains roughly constant but the hardening factor increases.

\begin{figure}
\centering
  \includegraphics{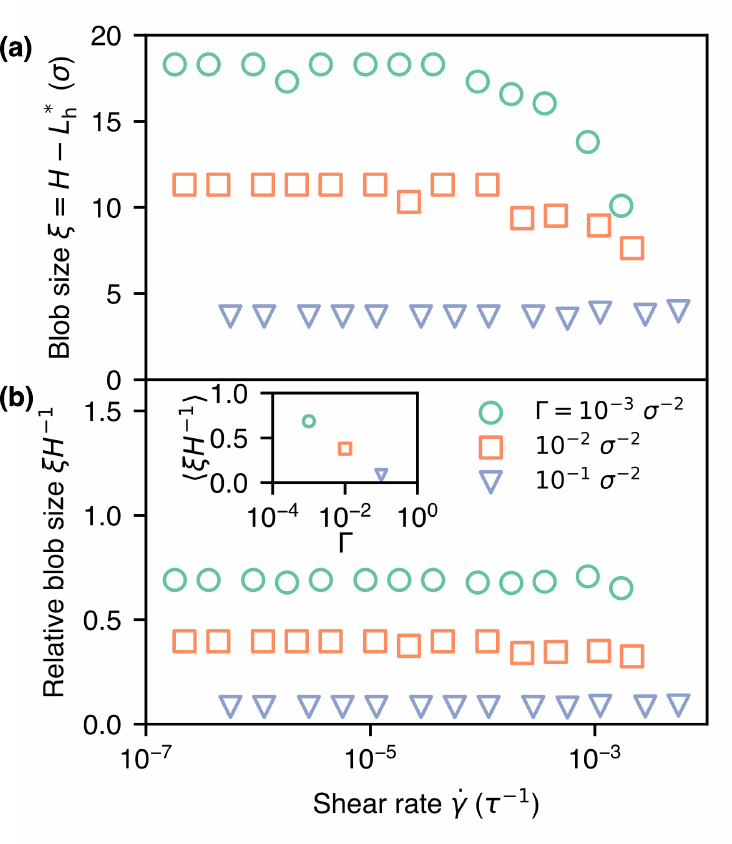}
  \caption{(a) Size of the outermost blob (correlation volume) as a function of grafting density $\Gamma$ and shear rate $\dot{\gamma}$ for systems with 60~beads/chain. (b) Blob size normalized by the shear rate-dependent brush height. Inset shows average values as a function of grafting density.}
  \label{fgr:blobsize}
\end{figure}

However, the theoretical model by Clement et al.~\cite{Clement2001EPL} states that the shear response is determined by the relaxation time of the brush's outermost correlation volume or blob. In Figure~\ref{fgr:blobsize}a we depict the blob's characteristic size $\xi$ as a function of grafting density and shear rate. We observe that while for our densest system the blob size is shear rate-independent over the full shear rate range, $\xi$ decreases at high shear rates for semi-dilute and dilute systems. For all systems and shear rates, the blob's size $\xi$ is proportional to the brush height $H$ (Fig.~\ref{fgr:blobsize}b).

\begin{figure}
\centering
  \includegraphics{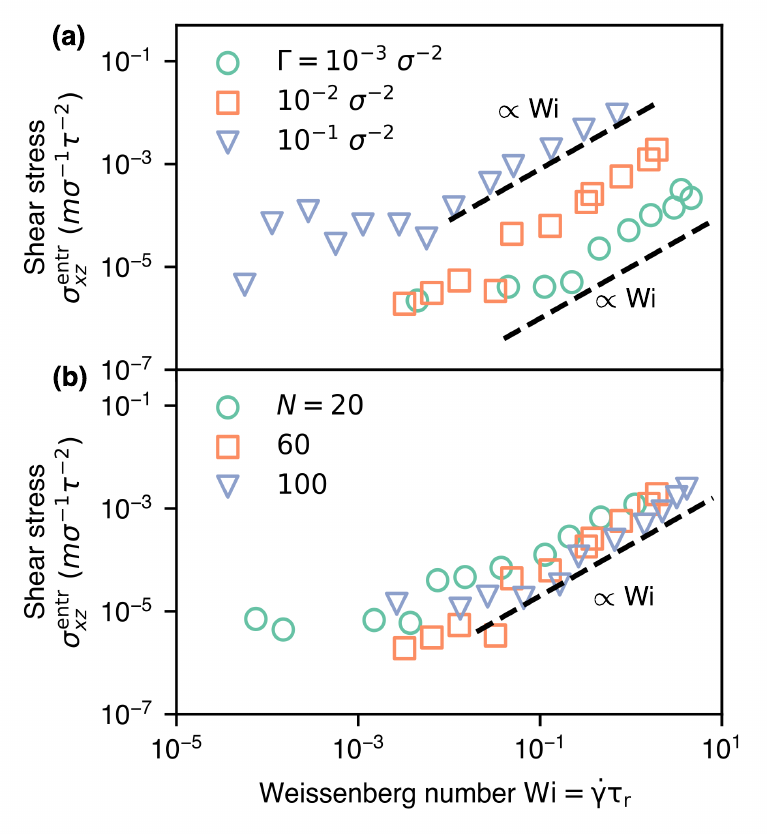}
  \caption{(a) Entropic shear stress as a function of grafting density $\Gamma$ for systems with 60~beads/chain. (b) Entropic shear stress as a function of the chain length for systems with $\Gamma=10^{-2}~\sigma^{-2}$.}
  \label{fgr:entropicstress}
  \end{figure}

We now analyze the entropic shear stress as a function of the Weissenberg number $\text{Wi}=\dot{\gamma}\tau_\text{r}$, where $\tau_\text{r}$ is the relaxation time of the blob. Given that the size of the outermost blob is shear-rate dependent, so is its relaxation time $\tau_\text{r}\approx\eta_\text{s}\xi^3/k_\text{B}T$. Figure~\ref{fgr:entropicstress}a shows that all systems studied here show two regimes. At low Wi,
%LP2JM: Move to discussion: the flow is too weak to substantially perturb chain conformations and
$\sigma^{\text{entr}}_{xz}$ shows little to no change with Wi. As Wi increases above a characteristic value of $\text{Wi}\approx 10^{-1}$,  a linear dependence of shear stress on Weissenberg number emerges.
We also observe that the shear stress increases with grafting density.
Both regimes exist independently of chain length, as evident from Fig.~\ref{fgr:entropicstress}b that shows the entropic stress as a function of Wi for chains of different length at the same grafting density.
Estimating the density for transition between mushroom and brushy regime yields $\Gamma^*~\approx~3.3~\cdot~10^{-2}~\sigma^{-2}$ for $N=20$ and $\Gamma^*~\approx~4~\cdot~10^{-3}~\sigma^{-2}$ for $N=100$.
This means the short chains are in the mushroom regime while the longer chains are already brushy, yet both behave in a similar fashion.
Crucially, the sublinear scaling regime disappears when using Wi rather than Wi$^*$ as the control parameter.

\subsection{Chain conformation}
\begin{figure*}
\centering
  \includegraphics{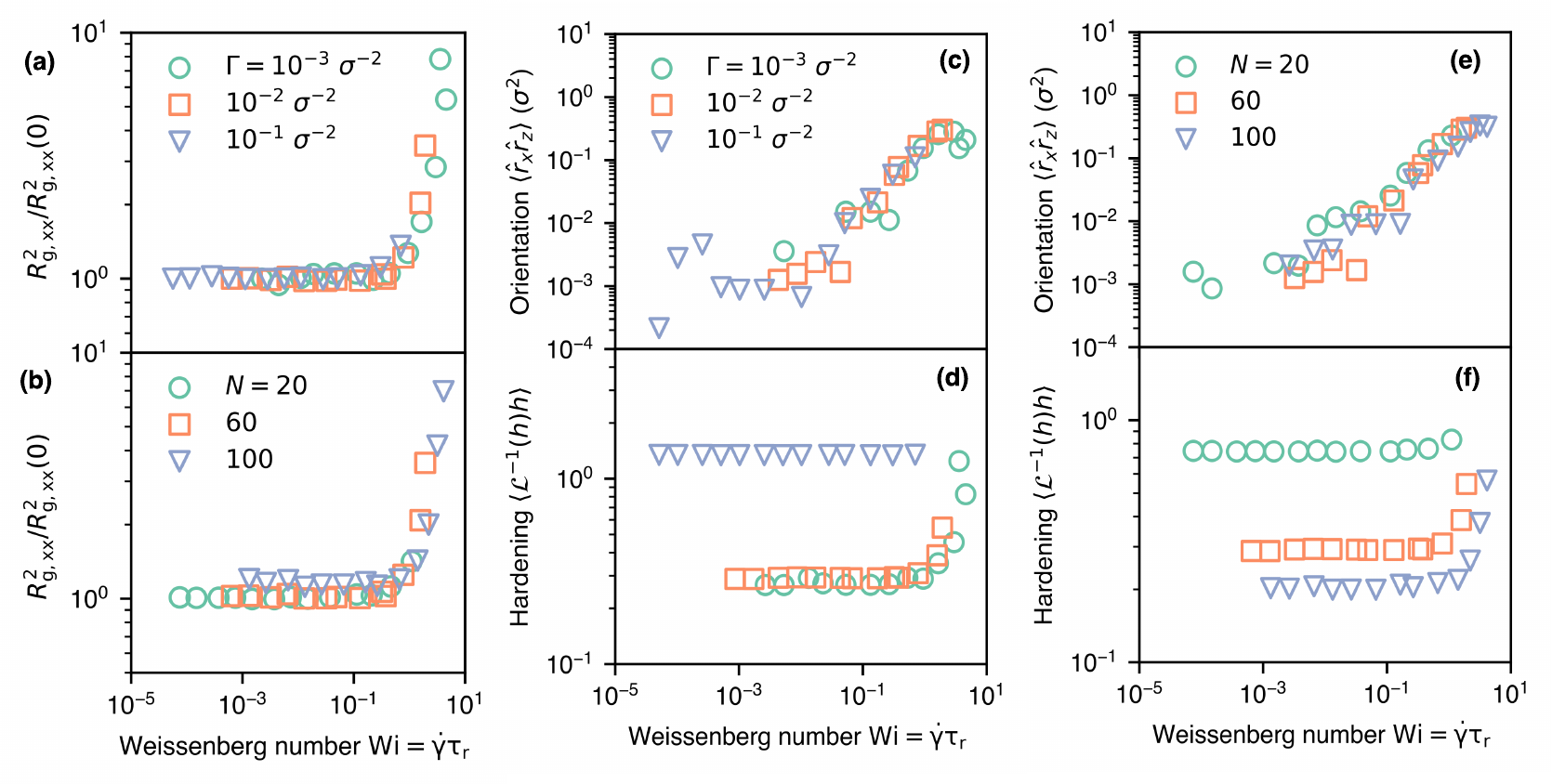}
  \caption{ Normalized gyration tensor in the direction of shear as a function of (a) grafting density $\Gamma$ (with $N=60~\mathrm{beads/chain}$) and (b) chain length $N$ (with $\Gamma=10^{-2}~\sigma^{-2}$). Orientation and hardening factors as a function of grafting density ((c) and (d)) and chain length ((e) and (f)).}
  \label{fgr:composite}
\end{figure*}
The question arises of how the shear flow affects the chain's conformations in the direction of shear. 
A common measure for conformation of a single chain is the gyration tensor,
\begin{equation}
    R^2_{\text{g},\alpha\beta}
    =
    \frac{1}{N}\sum_i
    (x_{i,\alpha} - x_{\text{CM},\alpha})
    (x_{i,\beta} - x_{\text{CM},\beta}),
\end{equation}
where $\vec{x}_i$ is the position of bead $i$ and $\vec{x}_\text{CM}$ is the center of mass of the chain.
Normalizing the gyration tensor in the direction of shear by its equilibrium value, we observe that the measurements collapse onto the same curve, despite changing grafting densities (Fig.~\ref{fgr:composite}a) or chain lengths (Fig.~\ref{fgr:composite}b). In both cases, the gyration tensor in the direction of shear remains constant up-to a threshold of $\text{Wi}\approx1$, at which point it starts increasing rapidly. Note that the Weissenberg number $\text{Wi}=\dot{\gamma}\tau_\text{r}$ seems to asymptotically approach a constant value for high shear rates, because an increase of $\dot{\gamma}$ is compensated by a decrease of $\tau_\text{r}$.  

A complementary view is offered by the orientation and hardening factors which comprise the entropic stress. In Figure~\ref{fgr:composite}c and e we observe that the orientation factors fall onto the same curve, being independent of Wi at low Wi and increasing linearly with Wi above approximately $\text{Wi}\approx10^{-1}$.
%At the highest Wi analyzed, the orientation factor seems to saturate at it's maximum value.
For the densest brush analyzed, the hardening factor is independent of Wi over the range analyzed. In contrast, our less dense systems show an increase of the hardening factor starting at approximately $\text{Wi}\approx1$. Furthermore, their hardening factor is lower in the Wi-independent regime. When varying the chain length, we observe that the hardening factor increases with decreasing chain length in the Wi-independent regime. Furthermore, systems with longer chains show a stronger increase in hardening factor at high Wi.

\begin{figure*}
\centering
  \includegraphics{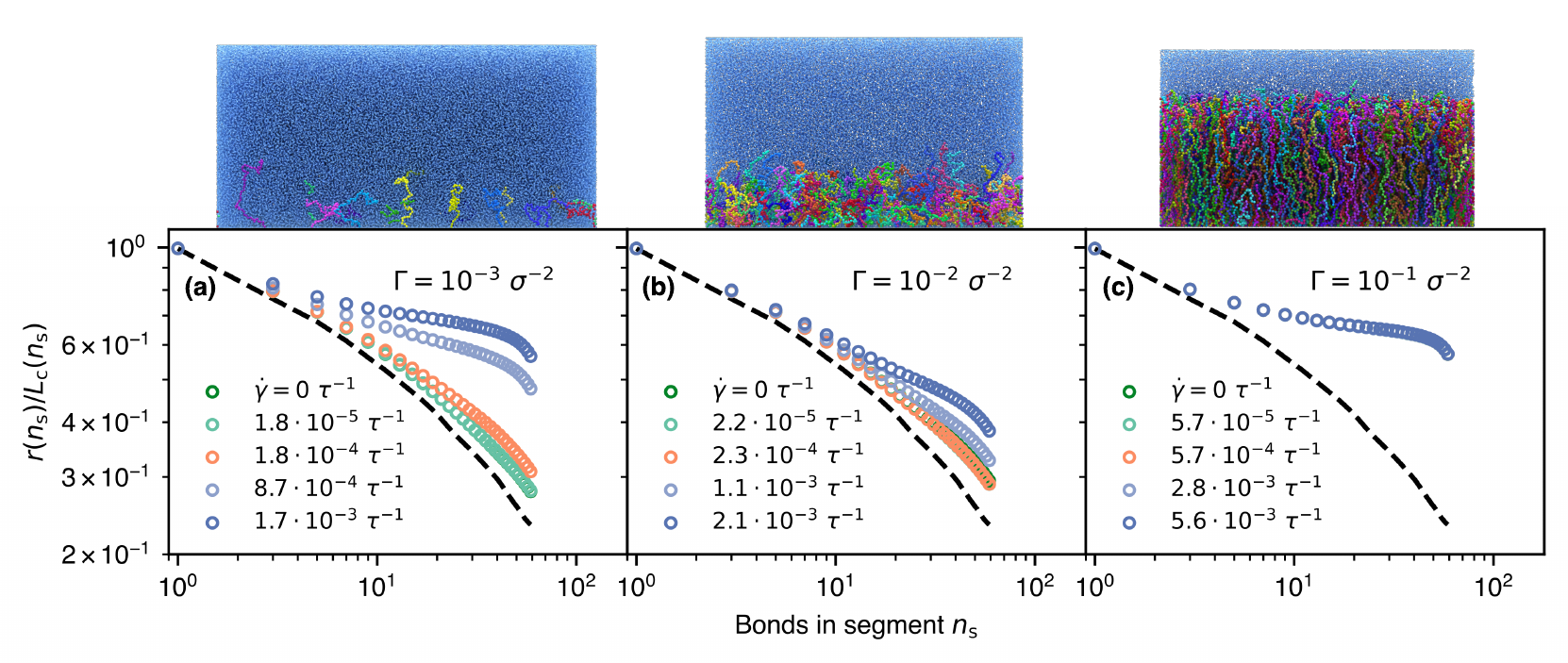}
  \caption{End-to-end distance $r(n_\text{s})$ of a chain fragments with $n_\text{s}$ bonds divided by their contour length $L_\text{c}=n_\text{s}b$ as a function of shear rate $\dot{\gamma}$ for systems with $N=60$~beads/chain. The panels (a), (b) and (c) show results for different grafting densities $\Gamma$. The first bead in every segment is bonded to the wall. Dashed lines represent results for a free chain with $N=60$~beads in dilute solution. A representative snapshot of the system in equilibrium ($\dot{\gamma}=0~\tau^{-1}$) is depicted for each grafting density.}
  \label{fgr:reeLc}
\end{figure*}

Finally, we investigate how the shear response within a chain changes as a function of distance to the wall. We do so by computing the end-to-end distance of chain fragments with $n_\text{s}$ bonds, normalized by the contour length $L_\text{c}=n_\text{s}b$ of said segment. The segments are chosen such that the first bead is tethered to the wall. % For dilute, undisturbed, real chains in equilibrium, $r/L_\text{c}\approx(n_\text{s}+1)^\nu/n_\text{s}$, with $\nu$ being Flory's exponent. 
Figure~\ref{fgr:reeLc} shows that for the lowest grafting density, the normalized end-to-end distance is reasonably close to results for a single, dilute chain at low shear rates. By comparison, as the grafting density increases, the normalized end-to-end distance in equilibrium deviates increasingly from the behavior of the single chain.    
At low shear rates, solvent flow does not affect chain conformations in all grafting densities analyzed.  As the shear rate increases, the chains elongate significantly in both dilute and semi-dilute systems. While the flow's effect can be observed close to the wall for the  dilute system, the point at which differences can be first noticed shifts further away from the wall (higher $n_\text{s}$) for the semi-dilute system. The dense system's morphology appears independent of shear rate in the range analyzed. 

%\jm{values dont make sense (too small for dilute and semi-dilute),probably because end is not necessarily at the edge of the brush} A similar procedure can be used to estimate the number of bonds within the brush's outermost blob. The end-to-end distance is computed starting from the last bead in the chain (i.e. the chain end not bonded to the wall) and normalized by the blob's size $\xi$. The relation $r(n_\text{s})/\xi \approx 1$ can be used to estimate the number of bonds within the outermost blob. For our systems with $N=60$~beads/chain (59~bonds), there are approximately 20~bonds within the outermost correlation volume for the dilute system, 15~bonds in the semi-dilute system and 7~bonds in the densest system.

\section{Discussion}
The grafting density's impact on the solvent's flow emerges from Fig.~\ref{fgr:grafted_chains}: At a grafting density close to the overlap concentration $\Gamma=1/\pi R_\mathrm{g}^2$, the solvent's  velocity in the direction of shear decays gradually within the brush until it reaches the wall. At a higher grafting density, a sharp transition occurs at the brush surface and the solvent's mean velocity is approximately zero within the brush. This is consistent with Milner's results showing that the hydrodynamic penetration length increases with decreasing grafting density~\cite{Milner1991M}. 

%It follows from the above considerations that in a semi-dilute or dilute system of grafted chains, the shear stress within the brush will be composed of the sum of the hydrodynamic and entropic stresses. Evidence is given by Fig.~\ref{fgr:walls}a, where the sum of the entropic stress and hydrodynamic stress from within the brush equals the system's shear stress measured at the walls. Note that the polymer's contribution to the shear stress is larger than the solvent's contribution. For dense brushes, the entropic stress is approximately equal to the system's stress because the shear rate is approximately zero inside the brush, hence the solvent's contribution to the shear stress is negligible. In short, the entropic stress is able to fully describe the system's shear response for systems of densely grafted chains and is a good approximation for semi-dilute systems.

It follows from the above considerations that in a semi-dilute or dilute system of grafted chains, the shear stress within the brush will be composed of the sum of the hydrodynamic and entropic stresses. However, our results show that the entropic shear stress is approximately equal to the system's stress and the solvent's contribution to the shear stress is negligible in both semi-dilute and dense systems. Given that the entropic stress is dependent on the brush's volume, Eq.~\eqref{eq:entropicstress}, our results validate the methodology used to determine the brush height.

When we compute the Weissenberg number with a constant relaxation time, the entropic shear stress is $\text{Wi}^*$-independent at low $\text{Wi}^*$, increases linearly thereafter and becomes sublinear at high $\text{Wi}^*$ (Fig.~\ref{fgr:relaxation}). %Chains are in quasi-equilibrium at low $\text{Wi}^*$. The linear regime coincides with an increase in the orientation factor, which stems from chains reorienting in the direction of shear. The sublinear regime coincides with an increase in the hardening factor, which correlates with chain elongation. 
%
%This means a change in size of the outermost correlation volume and thereby a decrease in the actual relaxation time.
%
We point out that similar scaling exponents have been found in the friction coefficients of monodisperse brushes~\cite{Galuschko2010Langmuir,Spirin2010EPJE}. An exponent of approximately 0.5 is also found in the Weissenberg number-dependent friction regime of self-mated hydrogels~\cite{Pitenis2014SM,Uruena2015Biotribology,Pitenis2018TRIL}. This regime has been attributed to both non-equilibrium polymer effects~\cite{Pitenis2014SM} and hydrodynamic lubrication~\cite{Dunn2014TRIL}. Our results indicate that the entropic stress of interfacial polymer chains might lie behind the observed frictional behavior of self-mated hydrogels. 

In a polymer brush, the shear response is defined by the time-scale of the relaxation of the brush's outermost blob or boundary layer~\cite{Clement2001EPL}, which depends on the blob's size, see Eq.~\eqref{eq:zimm}. In our densest system, the blob's size is independent of the shear rate (Fig.~\ref{fgr:blobsize}a). As a result, it is valid to approximate the system's behavior with a constant relaxation time. However, for the semi-dilute and dilute systems, the blob's size decreases significantly at high shear rates. Hence, it becomes more appropriate to use a shear-rate dependent relaxation time to compute the Weissenberg number. Given that the blobs size relative to the brush's height remains constant (Fig.~\ref{fgr:blobsize}b), the decrease in blob size $\xi$ can be attributed to a decrease of the brush's height. We posit that the increased polymer concentration leads to an increase in hydrodynamic screening, so that $L_\text{h}^*$ remains approximately constant. 

Using a shear rate-dependent relaxation time to compute the Weissenberg number Wi, we observe that the shear response can be split into two regimes: a Wi-independent regime and a regime in which the entropic stress rises linearly with Wi. This behavior emerges in both grafting density (Fig.~\ref{fgr:entropicstress}a) and chain length experiments (Fig.~\ref{fgr:entropicstress}b). The transition between both regimes at $\text{Wi}\approx10^{-1}$. Note that the sublinear regime at high Wi$^*$ is absent in the collapse with the the shear-rate dependent relaxation time.

We note that collapse of the curves observed for the gyration tensor in the direction of shear indicates that the choice of relaxation time $\tau_\text{r}$ captures the underlying polymer physics (Fig.~\ref{fgr:composite}a and b). The fact that systems with lower grafting densities are more susceptible to an increase of the gyration tensor at high Wi can be attributed to an increase in the solvent-polymer friction, which results from a bigger hydrodynamic penetration length (Fig.~\ref{fgr:composite}a). The force required to elongate the chain decreases with chain length, so system's with longer chains are more susceptible to an increase in the gyration tensor at high Wi   (Fig.~\ref{fgr:composite}b). 

We point out that for the gyration tensor in the shear direction, the Weissenberg number $\text{Wi}=\dot{\gamma}\tau_\text{r}$ becomes approximately constant at high shear rates. This entails that the relaxation time is inversely proportional to the shear rate in this regime. This is a consequence of the brush's outermost correlation volume $\xi$ decreasing in size with increasing shear rate. The hydrodynamic force acting on each correlation volume is equal to $6\pi\eta \dot{\gamma}\xi^2$~\cite{Harden1996PRE}, and hence increases with shear rate. This observation is supported by the fact that the gyration tensor in the shear direction continues to increase in this regime.  The outermost correlation volume decreasing in size with increasing force is reminiscent of Pincus' tension blobs~\cite{Pincus1976M}. Our results are in agreement with the assumption made in previous theoretical models that polymer brushes in shear flow can be described as strings of Pincus blobs~\cite{Rabin1990EPL,Barrat1992M,Harden1996PRE,Aubouy1996JdPII}.

By decomposing the entropic stress into the orientation and hardening factors (Fig.~\ref{fgr:composite}c-f), we confirm that the Wi-independent regime observed in the entropic shear stress at low Wi can be attributed to polymer chains relaxing faster than they are deformed by shear. The inflection point at which an increase in entropic stress is observed coincides with an increase in orientation factor. This can be interpreted as chains reorienting in the direction of shear. Finally, the increase in hardening factor coincides with the shear rate range, at which Wi becomes approximately constant.  It follows that an increase in chain elongation is accompanied by a change in size of the brush's outermost correlation volume. When using a constant relaxation time to compute the Weissenberg number (Fig.~\ref{fgr:relaxation}), the increase in hardening factor coincides with the sublinear regime. 

The hardening factor is higher for the densest system at low Wi because chains adopt extended conformations due to excluded volume interactions. At lower grafting densities, excluded volume interactions are not significant enough to affect chain conformations (Fig.~\ref{fgr:composite}d). The hardening factor decreases with increasing chain length due to the fact that the end-to-end distance approaches the contour length with decreasing chain length  (Fig.~\ref{fgr:composite}f). As a result, the extension ratio $h$ approaches unity as $N\rightarrow 0$. Unlike the hardening factor, the orientation factor is approximately equal in magnitude for all chain lengths considered (Fig.~\ref{fgr:composite}e). It follows that the collapse of the entropic shear stress observed in Fig.~\ref{fgr:entropicstress}b can be attributed to the number density in Eq.~\eqref{eq:entropicstress}.

Next, we turn to conformations within the chain. We attribute small deviations found in the normalized end-to-end distance of dilute systems of grafted chains and the behavior of a free, dilute single chain to long-range hydrodynamic interactions between the chains and to the influence of being connected to the wall (Fig.~\ref{fgr:reeLc}). As the grafting density increases and interactions amongst neighboring chains increase, the normalized end-to-end distance deviates further from the behavior of the single chain. Results for our densest system are characteristic for systems in which the chains are primarily in elongated, linear conformations.

The fact that the normalized end-to-end distance becomes susceptible to shear flow closer to the wall in the dilute system than in the semi-dilute system can be attributed to the larger solvent flow within the chains in the dilute system and resulting higher blob size $\xi$ (Fig.~\ref{fgr:reeLc}). Our results show that chain conformations become increasingly linear due to the influence of the shear flow. The lack of shear rate-dependency in the dense system can be attributed to the lack of shear flow penetration into the brush.

%The transition from monodisperse to polydisperse systems leads to the linear and sublinear dependencies on $\text{Wi}^*$ merging into a single, sublinear regime (Fig.~\ref{fgr:polydispersity}).  This can be attributed to the fact that long chains leave equilibrium at lower shear rates than short chains. Furthermore, long chains are more exposed to the flow and there is some evidence that in polydisperse brushes longer chains shield shorter chains from the flow~\cite{Vos2009Polymer}. 
%Note that it is unclear to us how to define a shear-rate dependent relaxation time $\tau_\text{r}$ and thereby the actual Weissenberg number Wi for the polydisperse systems.

%The shear response of polydisperse brushes is reminiscent of friction in self-mated hydrogels, which are highly polydisperse, chemically crosslinked polymer networks with a brushy surface~\cite{Itagaki2019M, Johnson2021EM,Gombert2019AMI,Gombert2020SM}. In the case of self-mated hydrogel friction, the $\text{Wi}^*$-independent friction regime has been attributed to thermal fluctuations relaxing chains faster than they are deformed by the effect of shear~\cite{Pitenis2014SM,Uruena2015Biotribology,Mees2022arXiv}, which is consistent with our findings. Above a threshold $\text{Wi}^*$, the friction coefficient rises with approximately $\text{Wi}^{*0.5}$~\cite{Pitenis2014SM,Uruena2015Biotribology,Pitenis2018TRIL}. 

\section{Conclusions}

In summary, our results confirm that the brush's outermost correlation volume determines the system's shear response. We further show that the size of the outermost correlation volume decreases at high shear rates. This is relevant for the definition of the Weissenberg number Wi in brushy systems, as the actual Wi depends nonlinearly on shear rate because the polymer's relaxation time depends on it. Furthermore, we have shown that the entropic shear stress can be used to explain the shear response of a system of grafted chains. At low Wi, chains are in quasi-equilibrium and the entropic shear stress is constant. This regime is followed by chains reorienting in the direction of shear and the entropic stress increases linearly with Wi. Finally, at high Wi chains elongate, which leads to a sublinear exponent if a constant relaxation time is used to compute the Weissenberg number.

\section{Appendix}
\subsection{Flory's characteristic ratio}
We compute Flory's characteristic ratio from polymer melt simulations with chains of length $N$. The systems were first equilibrated at a temperature of $T=0.6~\varepsilon$ and pressure of $P= 5\cdot10^{-4}~\mathrm{m}\sigma^{-1}\tau^{-2}$. This procedure produced systems with a monomer density of approximately $\rho=0.87~\sigma^{-3}$. A second equilibration at constant volume followed at an increased temperature of $T=2.0~\varepsilon$ to increase diffusivity. Finally, at $T=0.6~\varepsilon$ the end-to-end distance of each chain was sampled every $10^3$~timesteps for $10^5$~timesteps. Flory's characteristic ratio was computed for each system using the relation $C_n=\langle R^2 \rangle/nb^2$, where $n=N-1$ is the number of bonds in a chain~\cite{rubinstein2003}. 

\begin{figure}
\centering
  \includegraphics{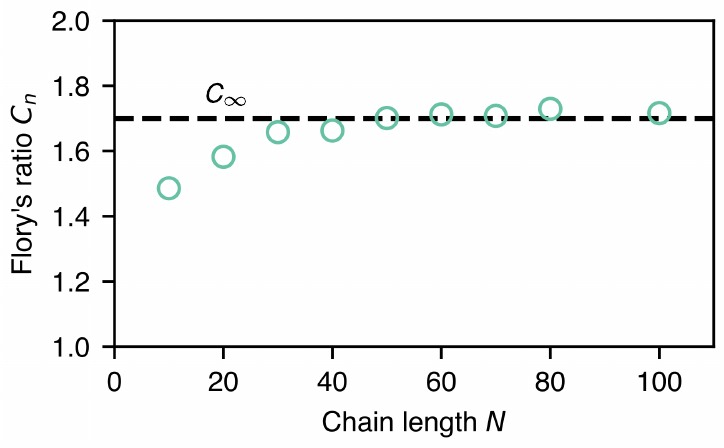}
  \caption{Flory's characteristic ratio as a function of chain length. Dashed line represents $C_\infty=1.7$.}
  \label{fgr:floryratio}
\end{figure}

We see in Figure~\ref{fgr:floryratio} that Flory's ratio increases rapidly with $N$ for $N < 30$~beads and saturates for higher $N$ at approximately 1.7. It follows that $C_\infty=1.7$.

\begin{acknowledgments}
We used \texttt{LAMMPS}~\cite{Plimpton1995JCompP} for all simulations and \texttt{Ovito~Pro}~\cite{Stukowski2009MSMSE} for visualization and post-processing. Computations were carried out on NEMO (University of Freiburg, DFG grant INST 39/963-1 FUGG) and HOREKA (project \textit{HydroFriction}). We thank the Deutsche Forschungsgemeinschaft for providing funding within grants PA~2023/2 and EXC-2193/1 –- 390951807.
\end{acknowledgments}

%\bibliography{references}% Produces the bibliography via BibTeX.


\end{document}